\documentclass[12pt,a4paper,twoside,english,thmsa,draft,titlepage]{article}

\usepackage[T1]{fontenc}
\usepackage[latin1]{inputenc}
\usepackage[french]{babel}

\usepackage{amsmath}
\usepackage{amsfonts}
\usepackage{amssymb}
\newcommand{\be}{\begin{equation}}
\newcommand{\ee}{\end{equation}}
\newcommand{\ba}{\begin{array}}
\newcommand{\ea}{\end{array}}
\newcommand{\bea}{\begin{eqnarray}}
\newcommand{\eea}{\end{eqnarray}}

\def\hbar{\not{\hbox{\kern-2.3pt $h$}}}
\def\psl{\not{\hbox{\kern-2.3pt $p$}}}
\def\Psl{\not{\hbox{\kern-2.3pt $P$}}}
\def\ksl{\not{\hbox{\kern-2.3pt $k$}}}
\def\qsl{\not{\hbox{\kern-2.3pt $q$}}}
\begin{document}
%
% \bibliographystyle{plain} %si on emploie Bibtex
% \bibliographystyle{unsrt} %si on emploie Bibtex
%
%%%%%%%%%%%%%%%%%%%%%%%%%%%%%%%%%%%%%%%%%%%%%%%%%%%%%%%%%
%
\begin{titlepage}
\today\hfill PAR-LPTHE 01/27
\begin{flushright} math-ph/0105050 \end{flushright}
\vskip 4cm
{\baselineskip 17pt
\begin{center}
{\bf TWO-DIMENSIONAL PARALEL TRANSPORT : \\ COMBINATORICS AND FUNCTORIALITY}
\end{center}
}
\vskip 5mm
\centerline{Romain Attal \footnote[1]{E-mail: attal@lpthe.jussieu.fr}}
\vskip 5mm
\centerline{{\em Laboratoire de Physique Th\'eorique et Hautes \'Energies}
     \footnote[2]{LPTHE tour 16\,/\,1$^{er}\!$ \'etage,
          Universit\'e P. et M. Curie, BP 126, 4 place Jussieu,
          F-75252 Paris Cedex 05 (France)}
}
\centerline{\em Universit\'es Pierre et Marie Curie (Paris 6) et Denis
Diderot (Paris 7)}
\centerline{\em Unit\'e associ\'ee au CNRS UMR 7589}
\vskip 1.5cm
{\bf Abstract:} We extend the usual notion of parallel transport along a
path to triangulated surfaces. In the one-dimensional case, each path in a base
space $ X $ is lifted into a fiber bundle with connection and this operation
defines a morphism between the fibers above the boundary points of the path.
In two dimensions, we define a fibered category over the space $ \widehat{\Pi } $
of paths in $ X $ and we associate to each homotopy of paths a functor between
the fibers above the two boundary paths of the swept surface. These ''sweeping
functors'' transport the information from the initial path to the final one.
We discuss the abelianisation problem and we conjecture that the smooth limit of the sweeping
functors, although not invariant under the action of the group of diffeomorphisms
of the swept surface, will provide a space of representation for this group.
\smallskip

{\bf PACS:} xxxx
% \centerline{\rule {3cm} {0.2pt}}
%\vfill
%\null\hfil\epsffile{.../LogoCNRS.ps}
%
\end{titlepage}
%%%%%%%%%%%%%%%%%%%%%%%%%%%%%%%%%%%%%%%%%%%%%%%%%

\section{Introduction and physical motivation}

\vspace{1.0cm}

In classical and quantum physics, point particles carry information (momentum,
spin, color, ... ) along their world-line and this evolution is obtained by lifting
a path $ (\gamma :[0,1]\rightarrow X) $, described by this particle in the
space-time manifold $ X $, into a vector bundle $ (E\rightarrow X) $ carrying a 
representation, $R$, of a group $G$. Then a connection on $ E $ provides a map 
$ (f_{\gamma }:F_{x}\rightarrow F_{y}) $ between the fibers $ F_{x} $ and $ F_{y} $ 
above the end points $ x=\gamma (0) $ and $ y=\gamma (1) $. 
$ f_{\gamma} $ is the parallel transport operator associated to
all these data. When the path is closed $(x=y)$, we obtain an automorphism of the 
fiber $ F_x $, the trace of which is invariant under the action of $G$. 
Since the work of K. Wilson \cite{Wilson}, we have a criterion for confinement 
of quarks in strongly coupled non-Abelian gauge theories : the basic observables are holonomy 
correlators, $W({\gamma })=\langle {\mathrm {tr}}(f_{\gamma})\cdot \cdot \cdot \rangle$. 
The asymptotics  $W({\gamma })\sim \exp (-\lambda A_{min}(\gamma))$, valid in the strongly 
coupled phase, reflects the collimation of flux lines and the concentration of energy on 
stringlike excitations sweeping a surface in space-time, with a string tension
$\sim \lambda$. Moreover, if we wish to define order/disorder operators in
Yang-Mills theory, analogous to those defined in two dimensions, we need a
general notion of two-dimensional parallel transport. In order to
extend the definition of holonomies to surfaces mapped into a manifold $ X $,
we will try to answer the following questions :\\

- Which fibered spaces are adapted to lift surfaces ?

- How to define a non-Abelian notion of two-dimensional parallel transport ? \\

It is known that gerbes (a kind of fibered space whose fibers are categories) are
the correct setting to answer these questions (\cite{MP}, \cite{Bry}, \cite{Gir}),
but we do not yet have a full understanding of these spaces when the symmetries
form a non-Abelian group, as in gauge theories.
\\ Let $ G $ be a locally compact, topological group.
This will be our local symmetry group.
We will take a polyhedron $ P $ as a base space in order to have a locally finite
number of degrees of freedom. We have in mind the triangulation of a surface
immersed in a manifold. Our building blocks will be the vertices, edges and
triangles of this polyhedron. If we want to define a discretized gauge theory
from these data, we can associate a copy of $ G $ to each vertex and an
element of $ G $ to each edge, and interpret this as a discretized version of
$ G $-bundle with connection. More generally, in order to describe non-trivial
bundles, we associate to each vertex a manifold having $ G $ as group of
automorphisms (the fiber is a $ G $-torsor) and to each edge a morphism
between the neighbouring fibers. These data define a functor from the category
$ \Pi  $, which has the vertices of $ P $ as objects and its edge-paths as
arrows, to the category $ G_{1} $ of right $ G $-torsors. These categories are
in fact groupoids since their arrows are isomorphisms. We extend this
construction one dimension higher using not only the vertices and edges but
also the oriented triangles of $ P $. With these building blocks we define a
groupoid $ \widehat{\Pi } $ whose objects are the edge-paths of $ P $ and
whose arrows are the simplicial homotopies between these edge-paths. Then we
can send $ \widehat{\Pi } $ in a suitable groupoid $ \mathcal{G} $ to
represent algebraically our triangulated surfaces. The group $ G $ acts on the
arrows of $ \mathcal{G} $ which are local degrees of freedom associated to the
triangles. Then we give a combinatorial definition of the operator which
represents our triangulated surface in $ \mathcal{G} $. This is not a
morphism, as was the case in one dimension, but a functor (between fibered
categories) which we call sweeping functor (SF). \\

In \S 2, we review some elementary concepts of category theory. We have tried
to provide concise definitions and a few examples to be as self-contained as
possible. In \S 3, we introduce the simplicial objects which are the building
blocks of the groupoids $ \Pi  $ and $ \widehat{\Pi } $. In \S 4,
we give a definition of simplicial $ G $-bundles. In \S 5, we extend this
definition to fibered categories and we define the sweeping functors as the
parallel transport operation in fibered categories. In \S 6, we illustrate these
concepts with a simple example : a 2-sphere triangulated by a tetrahedron and
swept in two different ways in order to compare the sweeping functors thus defined.
In \S 7, we discuss the abelianisation problem. A few perspectives are presented in \S 8. 
\vspace{1.0cm}

\section{Category vocabulary}

\vspace{1.0cm}

\subsection{Basic definitions}

In this section, we define the concepts of category, groupoid, monoidal
category and 2-category \cite{cwm}. When we compose maps in a category, it
is often easier to follow the arrows. Thus we will have
$ uvw\cdot \cdot \cdot =\cdot \cdot \cdot \circ w\circ v\circ u $.\\

\textbf{Definitions :}\\

a) A category $ \mathcal{C} $ consists in the following data :

- a class of objects $ O $ ;

- a class of arrows (or morphisms) $ A $ ;

- two maps $ s,t:A\rightarrow O $ (source and target) ;

- an associative composition of arrows $ ((u,v)\mapsto uv) $ defined whenever
$ t(u)=s(v) $ ;

- for each $ x\in O $, an identity arrow $ 1_{x} $ which is a left and
right neutral element for the composition.\\

b) A functor $ (f:\mathcal{C}\rightarrow \mathcal{C}^{\prime }) $ is a map
which sends each object $ x\in O $ to an object $ f(x)\in O^{\prime } $
and each arrow $ u\in A $ to an arrow $ f(u)\in A^{\prime } $ such that
$ f(uv)=f(u)f(v) $ whenever $ u $ and $ v $ are composable. If $ f $
reverses the arrows, i.e. $ f(uv)=f(v)f(u) $, it is called a contravariant
functor, or simply cofunctor.\\

c) A natural transformation $ (n:f\rightarrow g) $ between two functors
$ (f,g:\mathcal{C}\rightarrow \mathcal{C}^{\prime }) $
is a class of arrows $ (n_{x}:f(x)\rightarrow g(x)) $, one for each $ x\in O $.
The naturality of $ n $ means that for any arrow $ (u:x\rightarrow y) $
of $ \mathcal{C} $ we have the commuting square in $ \mathcal{C}^{\prime } $
:\\
 \[
f(u)\, n_{y}=n_{x}\, g(u)\]
\\
 The composition of arrows in $ \mathcal{C}^{\prime } $ induces a similar
composition of natural transformations
$ (f\stackrel{n}{\rightarrow }g\stackrel{m}{\rightarrow }h) $
with $ n_{x}m_{f(x)}=(nm)_{x} $ $ \forall x\in O $.
The functors $ (f:\mathcal{C}\rightarrow \mathcal{C}^{\prime }) $
and their natural transformations form a category noted
$ \mathcal{F}(\mathcal{C},\mathcal{C}^{\prime }) $.\\

d) A functor $ f\in \mathcal{F}(\mathcal{C},\mathcal{C}^{\prime }) $ is an
equivalence if there exists another functor
$ g\in \mathcal{F}(\mathcal{C}^{\prime },\mathcal{C}) $
and two natural transformations
$ (n:fg\rightarrow \mathrm{Id}_{\mathcal{C}}) $
and $ (n^{\prime }:gf\rightarrow \mathrm{Id}_{\mathcal{C}^{\prime }}) $
whose components are isomorphisms. \\

e) A groupoid is a category where all the arrows are isomorphisms.\\

f) A category $ \mathcal{C} $ is said monoidal if it is equipped with
a functor $ (\times :\mathcal{C}\times \mathcal{C}\rightarrow \mathcal{C}) $,
a unit object $ I $, and natural isomorphisms\\
\begin{eqnarray*}
a[A,B,C] & : & (AB)C\stackrel{\sim }{\longrightarrow }A(BC)\\
\ell _{A} & : & IA\stackrel{\sim }{\longrightarrow }A\\
r_{A} & : & AI\stackrel{\sim }{\longrightarrow }A
\end{eqnarray*}
\\
 called, respectively, associators, left unit isomorphisms and right unit isomorphisms,
and satisfying the pentagon coherence rule \\
\begin{eqnarray*}
 &  & \left( ((AB)C)D\stackrel{a[AB,C,D]}{\longrightarrow }(AB)(CD)
 \stackrel{a[A,B,CD]}{\longrightarrow }A(B(CD))\right) \\
 & = & \left( ((AB)C)D\stackrel{a[A,B,C]\times 1_{D}}{\longrightarrow }
 (A(BC))D\stackrel{a[A,BC,D]}{\longrightarrow }A((BC)D)\stackrel{1_{A}\times a[B,C,D]}
 {\longrightarrow }A(B(CD))\right)
\end{eqnarray*}
\\
 and the triangle coherence rule \\
 \[
\left( (AI)B\stackrel{a[A,I,B]}{\longrightarrow }A(IB)\stackrel{1_{A}\times \ell _{B}}
{\longrightarrow }AB\right) =\left( (AI)B\stackrel{r_{A}\times 1_{B}}{\longrightarrow }AB\right) \]
\\

g) A gr-category is a monoidal groupoid $ \mathcal{C} $ where
every object $ A $ has an inverse $ A^{*} $ : there exists an isomorphism
$ (\varepsilon _{A}:A^{*}A\stackrel{\sim }{\rightarrow }I) $ and this implies
the existence of an isomorphism $ (\eta _{A}:AA^{*}\stackrel{\sim }{\rightarrow }I) $.
Roughly speaking, a gr-category is a category endowed with a group-like structure
\cite{JS}\cite{Breen1}.\\

\textbf{Examples :}\\

i) The finite dimensional vector spaces over a fixed field $ \mathbb {F} $
and their $ \mathbb {F} $-linear maps form the category $ Vect_{\mathbb{F}} $.\\

ii) A topological space $ T $ equipped with a continuous, free and transitive
left (resp. right) action of a topological group $ G $ is called a left (resp.
right) $ G $-torsor. A morphism of left (resp. right) $ G $-torsors is
a continuous, $ G $--equivariant map, i.e. $ (f:T\rightarrow T^{\prime }) $
satisfies $ f(g\cdot t)=g\cdot f(t) $ (resp. $ f(t\cdot g)=f(t)\cdot g $)
$ \forall t\in T $ and $ \forall g\in G $. Such maps are in fact homeomorphisms
and these data form the category of left (resp. right) $ G $-torsors, denoted
$ G_{1}^{L} $ (resp. $ G_{1}^{R} $). We will drop the $ L $ or $ R $
superscript and consider only right $ G $-torsors.\\

iii) A mix of the two previous examples is provided by the category of finite
dimensional $ \mathbb {F} $-linear representations of $ G $, where the
morphisms are the $ G $-equivariant linear maps. This category is denoted
$Rep_{\mathbb{F}}(G) $. The tensor product of representation endows
$Rep_{\mathbb{F}}(G) $
with a structure of monoidal category. \\

iv) In \S\ 5, we will use the gr-category $ G^{\prime } $ whose objects are
the $ G $-bitorsors. These are the topological spaces on which the group
$ G $ acts on the left and on the right, freely and transitively. The inverse
$ K^{-1} $ of a bitorsor $ K $ is obtained by exchanging the left and
right actions of $ G $. The groupoid $ G_{1} $ is itself a torsor
under the action of $ G^{\prime } $. Indeed, multiplying on the right any
right $ G $-torsor $ T\in O(G_{1}) $ by a bitorsor $ K\in O(G^{\prime }) $
defines a right $ G $-torsor $ TK $ : \\

\begin{eqnarray*}
TK & = & \{(t,k)\in T\times K\}/\sim \\
(t,k) & \sim  & (t\cdot g,\, g^{-1}\cdot k){}\hspace {5mm}\forall g\in G
\end{eqnarray*}
\\
 The inverse operation is the right multiplication by $ K^{-1} $ \cite{bitors}.
\\

\subsection{2-categories}

In a set, the elements are only zero-dimensional objects without internal structure.
A category contains also one-dimensional elements (the arrows) which encode
the internal structure of its objects. It is natural to extend this notion one
step higher to two-dimensions. This is the task of 2-categories.\\

\textbf{Definition :}\\

A 2-category $ \mathcal{C} $ is defined by the following data :

- a class of objects $ O $ ;

- a class of 1-arrows $ (A_{1}) $ and a class of 2-arrows $ (A_{2}) $ ;

- a 1-source map $ s_{1}:A_{2}\rightarrow A_{1} $ and a 1-target map
$ t_{1}:A_{2}\rightarrow A_{1} $ ;

- a $ 0 $-source map $ s_{0}:A\rightarrow O $ and a $ 0 $-target map
$ t_{0}:A\rightarrow O $ satisfying\\
\begin{eqnarray*}
s_{0}t_{1} & = & s_{0}s_{1}\\
t_{0}t_{1} & = & t_{0}s_{1}
\end{eqnarray*}
\\

- a horizontal composition $ \times :A_{p}\times A_{p}\rightarrow A_{p} $
defined whenever the $ 0 $-source of the first $ p $-arrow is the $ 0 $-target
of the second one and for $ p=1 $ or $ 2 $ ;

- a vertical composition $ \circ :A_{2}\times A_{2}\rightarrow A_{2} $ defined
whenever the 1-source of the first 2-arrow is the 1-target of the second one ;

- a left and right identity 1-arrow $ (1_{x}:x\rightarrow x) $ for each object
$ x $ and a left and right identity 2-arrow $ (\varepsilon _{u}:u\rightarrow u) $
for each 1-arrow $ u $.\\

\noindent If $ x,y\in O $, $ A_{1}(x,y) $ will denote the class of 1-arrows
having $ x $ as source and $ y $ as target, and if $ u,v\in A_{1}(x,y) $,
$ A_{2}(u,v) $ will denote the class of 2-arrows having $ u $ as 1-source
and $ v $ as 1-target. To simplify the notations, we will sometimes write
$ \alpha \beta  $ for $ \beta \circ \alpha  $ and $ uv $ for $ v\times u $.
A 2-arrow $ \alpha \in A_{2}(u,v) $ is invertible when there exists $ \beta \in A_{2}(v,u) $
such that $ \alpha \beta =\varepsilon _{u} $ and $ \beta \alpha =\varepsilon _{v} $.
A 1-arrow $ u\in A_{1}(x,y) $ is invertible if there exists $ u^{\prime }\in A_{1}(y,x) $
and an invertible 2-arrow $ \alpha \in A_{2}(uu^{\prime },1_{x}) $. This
notion of invertibility {}''up to homotopy{}'' is adapted to structures having
automorphisms. All these data must satisfy several axioms : associativity (up
to homotopy) of the composition of 1-arrows, coherence of these homotopies,
... \cite{cwm}. When all the 2-arrows of $ \mathcal{C} $ are invertible
and all its 1-arrows are invertible up to homotopy, we say that $ \mathcal{C} $
is a 2-groupoid.\\

\textbf{Examples :}\\

i) The class of all small categories, with the functors as 1-arrows and the natural transformations
between these functors as 2-arrows, form the prototype of 2-categories.\\

ii) In \S 6, we will define the 2-groupoid $ \Pi_{2} $ whose objects
are the vertices of our polyhedron, whose arrows are its edge-paths and whose
2-arrows are equivalence classes of simplicial homotopies sweeping the same
oriented 2-cells in (possibly) different orders.\vspace{1.0cm}

\section{Simplicial objects and groupoids}

\vspace{1.0cm}

We give here some basic definitions about simplicial objects \cite{May} and
we introduce the notion of simplicial groupoid. Our approach to homotopical
algebra is very naive but uses groupoids as main tool and functoriality
as a guideline.\\

\subsection{Simplicial objects}

Simplicial objects are the building blocks of the simplest topological spaces
: simplicial complexes. This is the only type of spaces that a computer can
handle directly.\\

\textbf{Definitions :}\\

- A simplicial complex is a family of sets $ K=(K_{p})_{p\in {\mathbb{N}}} $ such
that for each $ p\geq 1 $, $ K_{p} $ is a set of subsets $ \sigma \subset K_{0} $
which contain exactly $ (p+1) $ elements and such that any non-empty subset
of $ \sigma  $ belongs to $ K_{q} $ for some $ q\leq p $. \\
- An element of $ K_{0} $ is called a vertex and an element of $ K_{p} $
is called a $ p $-simplex. $ K $ has pure dimension $ d $ if $ K_{d}\neq \emptyset  $,
$ K_{p}=\emptyset  $ for any $ p\geq d+1 $, and $ K_{0} $ is the union
of the $ d $-simplices of $ K_{d} $. \\
- The realization of $ K $, denoted $ |K| $, is the set of (barycentric
coordinate) functions $ (\alpha :K_{0}\rightarrow [0,1]) $ whose support,
$ \mathrm{Supp}(\alpha ) $, is a simplex and such that $ \sum _{v\in K_{0}}\alpha (v)=1 $.
$ |K| $ will be endowed with the weak topology (generated by the inverse
images of open sets via the $ \alpha  $'s). \\
- If $ X $ is a locally compact topological space, a triangulation of $ X $
is a homeomorphism $ (f:|K|\rightarrow X) $ and $ P=(X,K,f) $ is called
a polyhedron. \\
- The edges of $ P $ are the couples $ (x,y) $ where $ \{x,y\} $ is
a 1-simplex of $ P $. The degenerated edges are the couples $ (x,x) $
where $ x $ is a vertex of $ P $. A finite sequence of edges
$ ((x_{i},y_{i}))_{0\leq i\leq n} $
is composable if $ y_{i}=x_{i+1} $ for $ i=0,\cdot \cdot \cdot ,n-1 $. \\
- For all $ q\geq 0 $, the $ q $-skeleton of $ P $ is $ X^{q}=f(|K_{q}|) $,
where $ |K_{q}|=\{\alpha \in |K|,\mathrm{Supp}(\alpha )\in K_{q}\} $.\\

\subsection{Simplicial groupoids}

We define here the simplicial groupoid of a polyhedron as the category
built with the vertices and the edge-paths of this polyhedron.\\

\textbf{Definition :} Let $ \Pi  $ be the groupoid whose objects are
the vertices of $ P $ and whose arrows are the classes of finite sequences
$ ((x_{i},x_{i+1}))_{0\leq i\leq n-1} $ of composable edges, with respect
to the following equivalence relation :
$$(\cdot \cdot \cdot ,(x_{i-1},x),(x,x),(x,x_{i+2}),\cdot \cdot \cdot )
\sim (\cdot \cdot \cdot ,(x_{i-1},x),(x,x_{i+2}),\cdot \cdot \cdot ) $$
The composition of arrows is the concatenation of sequences. The sequences
of the form $ ((x,x),(x,x),(x,x),\cdot \cdot \cdot ) $ are identified to
$ ((x,x)) $ which is an identity arrow for this composition. The inverse
of $ ((x_{i},x_{i+1}))_{0\leq i\leq n-1} $ is the reversed path
$ ((x_{n-i},x_{n-i-1}))_{0\leq i\leq n-1} $.
$ \Pi  $ will be called the simplicial groupoid of $ P $.\\

\textbf{Definition :} Two edge-paths $ u=((x_{i},x_{i+1}))_{0\leq i\leq n-1} $
and $ v=((x_{i}^{\prime },x_{i+1}^{\prime }))_{0\leq i\leq n^{\prime }-1} $
are said $ X^{1} $-homotopic if they share the same end points and if there
exists a succession of cancellations or insertions of neighbouring opposite
edges which transform $ ((x_{i},x_{i+1}))_{0\leq i\leq n-1} $ into
$ ((x_{i}^{\prime },x_{i+1}^{\prime }))_{0\leq i\leq n^{\prime }-1} $.
$ u $ and $ v $ are said $ X^{2} $-homotopic if they share the same
end points and if $ u^{-1}v $ can be described by a loop which is contractible
in $ X^{2} $.\\

\textbf{Definition} : An oriented triangle of $ P $ is a $ q $-simplex
of $ P $ ($ q=0,1 $ or $ 2 $) with a couple of marked vertices $ (x,y) $
(its $ 0 $-source and $ 0 $-target) and a couple of marked edge-paths
$ (u,v) $ (its 1-source and 1-target) joining these vertices, such that :\\

(i) If $ q=2 $, $ uv^{-1} $ forms the boundary of a proper triangle.

(ii) If $ q=1 $ then $ u=v=(x,y) $ (flat triangle).

(iii) If $ q=0 $ then $ u=v=(x,x)=(y,y) $ (point triangle).\\

\noindent Consequently, the oriented triangles of $ P $ are of six kinds,
represented by the following pictures :\\

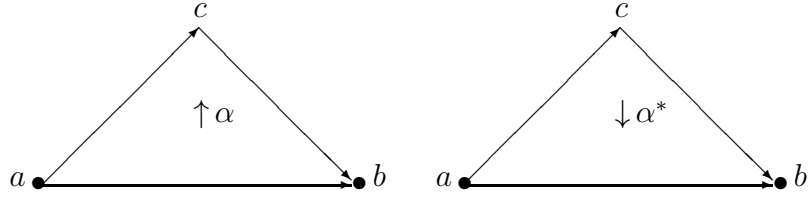
\unitlength=.7mm

\begin{picture}(100,50)(-40,-10)

\put(0,0){\vector(1,0){59}}
\put(0,0){\vector(1,1){30}}
\put(30,30){\vector(1,-1){29}}
\put(-2,-1){$\bullet$}
\put(59,-1){$\bullet$}
\put(-6,0){$a$}
\put(63,0){$b$}
\put(29,32){$c$}
\put(29,12){$\uparrow$}
\put(33,12){$\alpha$}

\put(80,0){\vector(1,0){59}}
\put(80,0){\vector(1,1){30}}
\put(110,30){\vector(1,-1){29}}
\put(79,-1){$\bullet$}
\put(139,-1){$\bullet$}
\put(75,0){$a$}
\put(143,0){$b$}
\put(109,32){$c$}
\put(109,12){$\downarrow$}
\put(113,12){$\alpha^{*}$}

\end{picture}\\

\noindent In these first two cases, the two marked points, $ a $ and $ b $,
are distinct. The left cell $ (\alpha :(ab)\rightarrow (ac,cb)) $ is the
unique oriented triangle going from the edge $ u=(ab) $ to the edge-path
$ v=(ac,cb) $ and the right cell $ (\alpha ^{*}:(ac,cb)\rightarrow (ab)) $
is the reversed triangle. We will note $ (acb) $ the $ \alpha  $-cells
and $ (acb)^{*} $ the (reversed) $ \alpha ^{*} $-cells.\\

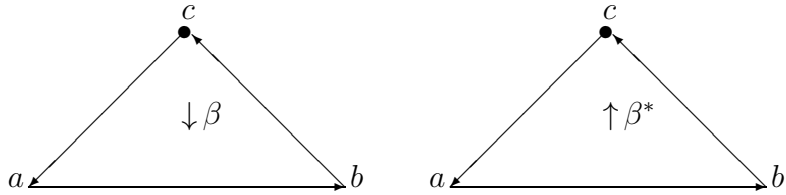
\begin{picture}(100,50)(-40,-10)

\put(0,0){\vector(1,0){60}}
\put(29,29){\vector(-1,-1){29}}
\put(60,0){\vector(-1,1){29}}
\put(28,28){$\bullet$}
\put(-4,0){$a$}
\put(61,0){$b$}
\put(29,32){$c$}
\put(29,12){$\downarrow$}
\put(33,12){$\beta$}

\put(80,0){\vector(1,0){60}}
\put(140,0){\vector(-1,1){29}}
\put(109,29){\vector(-1,-1){29}}
\put(108,28){$\bullet$}
\put(76,0){$a$}
\put(141,0){$b$}
\put(109,32){$c$}
\put(109,12){$\uparrow$}
\put(113,12){$\beta^{*}$}

\end{picture}\\

\noindent In these two other cases, the two marked points are identical
$ (x=y=c) $
and one of the two boundary edge-paths is reduced to the degenerated edge,
$ u=(cc)=1_{c} $,
while the other one, $ v=(ca,ab,bc) $, describes the whole boundary of the
2-simplex. The cell $ (\beta :(cc)\rightarrow (ca,ab,bc)) $ is the unique
oriented triangle going from the trivial edge $ (cc) $ to the edge-path
$ (ca,ab,bc) $
and $ (\beta ^{*}:(ca,ab,bc)\rightarrow (cc)) $ is the reversed triangle.
We will note $ (cabc) $ the $ \beta  $-cells and $ (cabc)^{*} $ the
(reversed) $ \beta ^{*} $-cells.\\

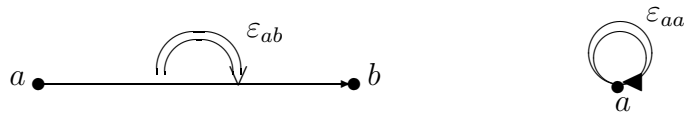
\begin{picture}(100,50)(-40,-15)

\put(0,5){\vector(1,0){59}}
\put(-2,3.5){$\bullet$}
\put(58,3.5){$\bullet$}
\put(-6,5){$a$}
\put(62,5){$b$}

\put(108,3){$\bullet$}
\put(109,0){$a$}
\put(110,10){\circle{10.5}}
\put(110,11){\circle{12.5}}

\put(110,4){$\blacktriangleleft$}

\put(115,17){$\varepsilon_{aa}$}

\put(30,7){\oval(16,16)[t]}
\put(30,7){\oval(13,13)[t]}
\put(39,14){$\varepsilon_{ab}$}

\put(35.5,5){$\vee$}

\end{picture}\\

\noindent The last two cases (degenerated triangles) are the identity arrows
and correspond, respectively, to $ q=1 $ (left) and $ q=0 $ (right).\\

We now define from $ \Pi  $ another groupoid $ \widehat{\Pi } $
which is the simplicial groupoid of the space of edge-paths in $ P $.
The objects of $ \widehat{\Pi } $ are the arrows of $ \Pi  $ and the arrows
of $ \widehat{\Pi } $ are defined by generators and relations. \\

\textbf{Definition :} The elementary arrows of $ \widehat{\Pi } $ are the
couples $ (\gamma ,\gamma ^{\prime }) $ of edge-paths such that :\\

(i) $ \gamma  $ and $ \gamma ^{\prime } $ have the same 0-source and 0-target
;

(ii) there exists an oriented triangle $ \sigma  $ such that $ \gamma  $
is $ X^{1} $-homotopic to $ s_{1}(\sigma ) $ and $ \gamma ^{\prime } $
is $ X^{1} $-homotopic to

\hspace{5mm}$ t_{1}(\sigma ) $, both relatively to the subset
$ \{s_{0}(\sigma ),t_{0}(\sigma )\} $,
i.e. these vertices staying fixed.\\

\noindent In the six cases, the $ p $-source and $ p $-target ($ p=0 $
or $ 1 $) of an elementary arrow are well defined. Since we are building
a groupoid, the elementary arrows must also be invertible. Thus we impose
the following relations between our generators :\\
\begin{eqnarray*}
(acb)^{*}\circ (acb) & = & \varepsilon _{ab}\\
(acb)\circ (acb)^{*} & = & \varepsilon _{(ac,cb)}\\
(cabc)^{*}\circ (cabc) & = & \varepsilon _{cc}\\
(cabc)\circ (cabc)^{*} & = & \varepsilon _{(ca,ab,bc)}
\end{eqnarray*}
\\

\noindent \textbf{Definition} : The arrows of $ \widehat{\Pi } $ are the
finite sequences $ \Gamma =((\gamma _{i},\gamma _{i+1}))_{0\leq i\leq n-1} $
of elementary arrows sharing the same 0-source and 0-target. The $ p $-source
and $ p $-target of $ \Gamma  $ are defined by\\
\begin{eqnarray*}
s_{0}(\Gamma ) & = & s(\gamma _{i})\\
t_{0}(\Gamma ) & = & t(\gamma _{i})\\
s_{1}(\Gamma ) & = & \gamma _{0}\\
t_{1}(\Gamma ) & = & \gamma _{n}
\end{eqnarray*}
\\
 If $ \Gamma =((\gamma _{i},\gamma _{i+1}))_{0\leq i\leq n-1} $ and
$ \Gamma ^{\prime }=((\gamma _{i}^{\prime },\gamma _{i+1}^{\prime }))_{0\leq i\leq n^{\prime }-1} $
satisfy $ \gamma _{n}=\gamma _{0}^{\prime } $, i.e. $ t_{1}(\Gamma )=s_{1}(\Gamma ^{\prime }) $,
their vertical composite is the family
$ \Gamma \Gamma ^{\prime }=((\gamma _{0},\gamma _{1}),\cdot \cdot \cdot
,(\gamma _{n^{\prime }-1}^{\prime },\gamma_{n^{\prime}}^{\prime })) $
obtained by concatenation.\\

The representations of $ \widehat{\Pi } $ will
provide us with the right notion of two-dimensional parallel transport. But
we first need a simplicial analogue of $ G $-bundles with connection, which
we introduce in the next section. \\

In order to parametrize our edge-paths and simplicial homotopies, we need a
simplicial analogue of the interval $ [0,1] $. This r\^{o}le is played by
the simplicial category $ \Delta $ whose objects are the finite groupoids 
$ \Delta_p $ given by\\

\begin{eqnarray*}
O(\Delta_p) & = & \{0,\cdots ,p-1 \}\\
A(i,j) & = & \{(i,j)\}{\hspace {10mm}}\forall {\hspace {3mm}}i,j\in \{0,\cdots ,p\}
\end{eqnarray*}
\\

The composition of arrows in $ \Delta_p $ is $ (i,j)(j,k)=(i,k) $ and the
identity arrows are the $ (i,i) $'s. The arrows of $ \Delta $ are
generated by composition of the shift maps \\
\begin{eqnarray*}
s_{pk}:\Delta_{p} & \rightarrow  & \Delta_{p+1}\hspace {33.5mm}\widetilde{s}_{p\ell }:\Delta_{p}\rightarrow \, \, \Delta_{p-1}\\
i & \mapsto  & i{\hspace {14mm}}(i\leq k)\hspace {1.0in}i\, \, \mapsto \, \, i{\hspace {20mm}}(i\leq \ell )\\
i & \mapsto  & i+1{\hspace {8mm}}(i>k)\hspace {1.0in}i\, \, \mapsto \, \, i-1{\hspace {14mm}}(i>\ell )
\end{eqnarray*}
\\
A symmetric monoidal structure is defined on
$ \Delta $ by $ \Delta_{p}+\Delta_{q}=\Delta_{p+q} $
and $ I=\Delta_{0}=\emptyset $.\\

\begin{eqnarray*}
A_{\Delta}(\Delta_{p},\Delta_{p+1}) & = & \{s_{pk}\, /\, k=0,\cdots ,p-1\}\\
A_{\Delta}(\Delta_{p},\Delta_{p-1}) & = & \{\widetilde{s}_{p\ell }\, /\, \ell =0,\cdots ,p-1\}
\end{eqnarray*}
\\
 An edge-path in $ P $ is now interpreted as a functor $ (\gamma :\Delta_{p}\rightarrow \Pi ) $, $ p $ being the number of edges of $ \gamma  $ or its length. $ \widehat{\Pi } $
is therefore the category of functors $ (\Delta_{p}\rightarrow \Pi ) $ :\\

 $$ O(\widehat{\Pi })=\bigcup _{p\in {\mathbb{N}}}O(\mathcal{F}(\Delta_{p},\Pi )) $$
\\
 The arrows of $ \widehat{\Pi } $ are generated by the natural transformations
between these functors (this supposes a common $ \Delta_{p} $ as source) and by
the homotopies connecting edge-paths of different lenghts.\vspace{1.0cm}

\section{Simplicial $ G $-bundles}

\vspace{1.0cm}

So far, we have encoded some topological information about the polyhedron $ P $
in the simplicial groupoids $ \Pi  $ and $ \widehat{\Pi } $. In
order to compute anything, we need an algebraic representation of these spaces.
We define below simplicial fiber bundles with connection over our polyhedron
$ P $ as representations of $ \Pi  $. We have already defined the groupoid
$ G_{1} $ of (left) $ G $-torsors. The latter are topological manifolds
which look like $ G $ but without marked point.\\

\textbf{Definition} :\ {}A simplicial $ G $-bundle with connection over the
polyhedron $ P $ is a functor $ (f:\Pi \rightarrow G_{1}) $ from the simplicial
groupoid $ \Pi  $ to the groupoid $ G_{1} $ of $ G $-torsors.
\\
The usual definition of a simplicial G-bundle refers to a simplical object in the 
category of G-bundles over some space. These objects come without connection.
We have used the same terminology hoping there will be no confusion.

\noindent Such a functor $ (f:\Pi \rightarrow G_{1}) $ associates to each vertex 
$ a\in X^{0} $ a $ G $-torsor $ f_{a} $ and to any edge-path $ \gamma  $ a morphism 
of $ G $-torsors $ (f_{\gamma }:f_{a}\rightarrow f_{b}) $. 
If $ \gamma  $ is an edge $ (ab) $,
we will write it $ f_{ab} $. When $ \gamma  $ is a loop based at $ a $,
$ f_{\gamma } $ is an automorphism of $ f_{a} $, i.e. an element of $ G $.
This is the holonomy of $ f $ along $ \gamma  $. From now on, we will
drop the epithet {}''simplicial{}'' which will be understood.\\

\textbf{Definition} : Let\ {}$ (f:\Pi \rightarrow G_{1}) $ be a $ G $-bundle
with connection over $ P $ and $ R=(V,\rho ) $ be a $ {F} $-linear
representation of $ G $. The vector bundle with linear connection associated
to $ f $ and $ R $ is the functor $ f_{R} $ defined by\\
\begin{eqnarray*}
f_{R}:\Pi  & \rightarrow  & (G_{1}\times {{Re}}\mathbf{p}_{{F}}(G))/\sim \\
x & \mapsto  & f_{R}(x)=(f_{x}\times V)/\sim \\
xy & \mapsto  & f_{R}(xy)=(f_{xy},\mathrm{Id}_{V})
\end{eqnarray*}
\\
 with the usual equivalence relation :\\
 \[
(t\cdot g,v)\sim (t,\rho (g)v)\]
\\
 A morphism of $ G $-bundles is a natural transformation $ (n:f\rightarrow g) $
between the underlying functors i.e. a morphism of $ G $-torsors
$ (n_{a}:f_{a}\rightarrow g_{a}) $
for each $ a\in X^{0} $, such that $ \forall a,b\in X^{0} $\\
 \[
f_{ab}\, n_{b}=n_{a}\, g_{ab}\]
\\
 Two $ G $-bundles $ f $ and $ g $ are isomorphic when there exists
an invertible natural transformation $ (n:f\rightarrow g) $. This implies
that $ f $ and $ g $ have the same holonomies. The automorphisms of a
$ G $-bundle $ f $ (its invertible natural transformations) form the gauge
group\\
 \[
\mathrm{Aut}(f)=\mathcal{F}(X^{0},G)\]
\\
 and the $ G $-bundles over $ P $, together with their natural transformations,
form the category $ \mathbf{Bun}_{G}(P)=\mathcal{F}(\Pi ,G_{1}) $.

\section{Fibered categories and sweeping functors}

\vspace{1.0cm}

We now extend the notion of $ G $-bundle with connection one step higher
on the $ n $-category ladder. Our approach can be related to the notion of
gerbe as developed in (\cite{ast}, \cite{Bry}, \cite{Gir}) and to that of combinatorial stack
\cite{Kap}. Although less general, the discretized versions are easier to visualize
than their sheafified parents.\\

\textbf{Definition} : A (simplicial) gerbe is a functor
$ \left( \varphi :\mathcal{B}\rightarrow \mathcal{G}\right)  $
from a base groupoid $ \mathcal{B} $ to a groupoid $ \mathcal{G} $
whose objects are themselves groupoids. The sections of the stack $ \varphi  $
are the functors $ (\sigma :\mathcal{B}\rightarrow \varphi (\mathcal{B})) $
such that for all $ x,y\in O(\mathcal{B}) $, $ \sigma (x)\in O(\varphi (x)) $
and if $ (u:x\rightarrow y) $ is an arrow of $ \mathcal{B} $,
$ (\sigma (u):\sigma (x)\rightarrow \sigma (y)) $
is an arrow of the sub-groupoid $ \varphi (\mathcal{B}) $.\\

From now on, we will focus on the following particular case. Let $ \mathcal{B}=\Pi  $
and let $ \mathcal{G}=\{G_{1}\} $ be the groupoid with $ G_{1} $
as the only object and its automorphisms as arrows. (These automorphisms are
the endofunctors obtained by multiplication on the right by a $ G $-bitorsor
$ K\in O(G^{\prime }) $). For each edge-path $ \gamma  $, $ \Pi _{\gamma } $
denotes the sub-groupoid of $ \Pi  $ generated by the vertices and
edges of $ \gamma  $, $ \varphi _{\gamma } $ is the restriction of $ \varphi  $
to $ \Pi _{\gamma } $, and $ S_{\gamma } $ is the category of sections
of $ \varphi _{\gamma } $. We define $ S $ as the fibered category over
$ \widehat{\Pi } $, whose fiber at $ \gamma  $ is the category $ S_{\gamma } $
and whose arrows between the fibers $ S_{\gamma } $ and $ S_{\gamma ^{\prime }} $
above two $ X^{2} $-homotopic paths $ \gamma  $ and $ \gamma ^{\prime } $
will be our sweeping functors. Since $ \varphi _{\gamma } $ and
$ \varphi _{\gamma ^{\prime }} $
are stacks above $ \gamma  $ and $ \gamma ^{\prime } $, a functor
$ \left( \psi :S_{\gamma }\rightarrow S_{\gamma ^{\prime }}\right)  $
must send a section $ \sigma  $ of $ \varphi _{\gamma } $ to a section
$ \psi (\sigma ) $ of $ \varphi _{\gamma ^{\prime }} $ and an arrow
$ (u:\sigma \rightarrow \tau ) $
between sections $ \sigma ,\tau \in O(S_{\gamma }) $ to an arrow
$ (\psi (u):\psi (\sigma )\rightarrow \psi (\tau )) $
between the image sections $ \psi (\sigma ),\psi (\tau )\in O(\varphi _{\gamma ^{\prime }}) $.
Since the arrows of $ \widehat{\Pi } $ were obtained by generators and relations,
it is sufficient to define $ \mathcal{F}\left( S_{\gamma },S_{\gamma ^{\prime }}\right)  $
for neighbouring paths forming an edge $ (\gamma ,\gamma ^{\prime }) $ of
$ \widehat{\Pi } $. So let $ \alpha =(acb) $ be an oriented triangle of
$ P $. Let $ (\sigma _{a},\sigma _{b}) $ be a section of $ (\varphi _{a},\varphi _{b}) $,
and $ (\sigma _{ab}:\varphi _{ab}(\sigma _{a})\rightarrow \sigma _{b}) $
be an arrow between these objects. The image
$ S_{\alpha }(\sigma _{a},\sigma _{b})=(\sigma _{a}^{\prime },\sigma _{c}^{\prime },\sigma _{b}^{\prime }) $
is the section of $ (\varphi _{a},\varphi _{b},\varphi _{c}) $ defined by
: \\
\begin{eqnarray}
\sigma _{a}^{\prime } & = & \sigma _{a}\nonumber \\
\sigma _{c}^{\prime } & = & \varphi _{ac}(\sigma _{a})\\
\sigma _{b}^{\prime } & = & \sigma _{b}\nonumber
\end{eqnarray}
\\
 and these objects are connected by the arrows \\
\begin{equation}
\sigma _{ac}^{\prime }=1_{\sigma _{c}}:\sigma _{c}\rightarrow \sigma _{c}
\end{equation}
\begin{equation}
\sigma _{cb}^{\prime }=\sigma _{ab}\circ (\varphi _{acb}(\sigma _{a}))^{-1}:
\varphi _{cb}(\sigma _{c}^{\prime })\rightarrow \sigma _{b}^{\prime }
\end{equation}
\\
 The inverse move associates to the arrows $ (\tau _{ac},\tau _{cb}) $ the
arrow $ (\tau _{ab}^{\prime }:\varphi _{ab}(\tau _{a})\rightarrow \tau _{b}) $
defined by \\
\begin{equation}
\tau _{ab}^{\prime }=\tau _{cb}\circ \varphi _{cb}(\tau _{ac})\circ \varphi _{acb}(\tau _{a})
\end{equation}
\\

Thus, to each oriented triangle $ \alpha =((acb):(ab)\rightarrow (ac,cb)) $
corresponds a functor $ (S_{\alpha }:S_{(ab)}\rightarrow S_{(ac,cb)}) $,
and to its inverse $ \alpha ^{*} $ corresponds $ S_{\alpha ^{*}}=(S_{\alpha })^{-1} $
:\\

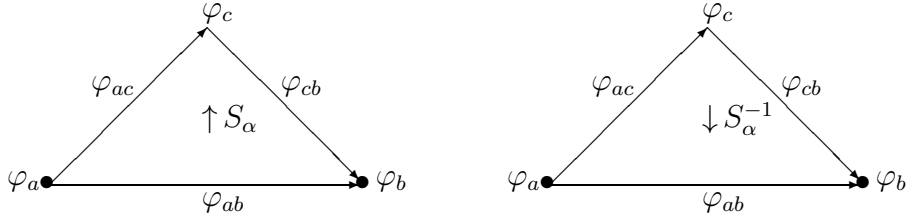
\begin{picture}(100,60)(-30,-10)

\put(0,5){\vector(1,0){59}}
\put(0,5){\vector(1,1){30}}
\put(30,35){\vector(1,-1){29}}
\put(-2,4){$\bullet$}
\put(58,4){$\bullet$}
\put(-8,5){$\varphi_{a}$}
\put(62,5){$\varphi_{b}$}
\put(29,37){$\varphi_{c}$}
\put(29,16){$\uparrow$}
\put(33,16){$S_{\alpha}$}
\put(29,1){$\varphi_{ab}$}
\put(8,23){$\varphi_{ac}$}
\put(44,23){$\varphi_{cb}$}

\put(95,5){\vector(1,0){59}}
\put(95,5){\vector(1,1){30}}
\put(125,35){\vector(1,-1){29}}
\put(93,4){$\bullet$}
\put(153,4){$\bullet$}
\put(87,5){$\varphi_{a}$}
\put(157,5){$\varphi_{b}$}
\put(124,37){$\varphi_{c}$}
\put(124,16){$\downarrow$}
\put(128,16){$S_{\alpha}^{-1}$}
\put(124,1){$\varphi_{ab}$}
\put(103,23){$\varphi_{ac}$}
\put(139,23){$\varphi_{cb}$}

\end{picture}\\

\noindent Similarly, to each oriented triangle $ \beta =((cabc):(cc)\rightarrow (ca,ab,bc)) $
corresponds a functor $ (S_{\beta }:S_{(cc)}\rightarrow S_{(ca,ab,bc)}) $,
and to its inverse $ \beta ^{*} $ corresponds $ S_{\beta ^{*}}=(S_{\beta })^{-1} $
:\\

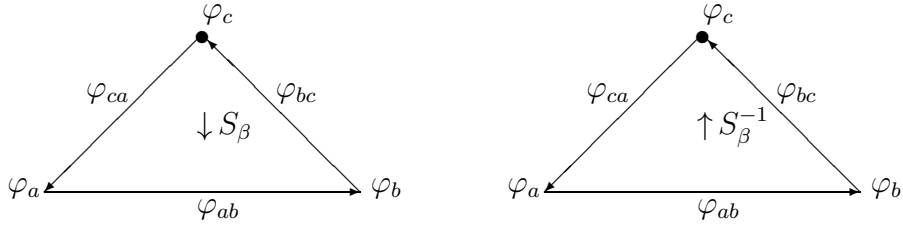
\begin{picture}(100,60)(-30,-10)

\put(0,5){\vector(1,0){60}}
\put(29,34){\vector(-1,-1){29}}
\put(60,5){\vector(-1,1){29}}
\put(-7,5){$\varphi_{a}$}
\put(62,5){$\varphi_{b}$}
\put(30,38){$\varphi_{c}$}
\put(28.5,33){$\bullet$}
\put(29,16){$\downarrow$}
\put(33,16){$S_{\beta}$}
\put(29,1){$\varphi_{ab}$}
\put(8,23){$\varphi_{ca}$}
\put(44,23){$\varphi_{bc}$}

\put(95,5){\vector(1,0){60}}
\put(124,34){\vector(-1,-1){29}}
\put(155,5){\vector(-1,1){29}}
\put(88,5){$\varphi_{a}$}
\put(157,5){$\varphi_{b}$}
\put(125,38){$\varphi_{c}$}
\put(123.5,33){$\bullet$}
\put(124,16){$\uparrow$}
\put(128,16){$S_{\beta}^{-1}$}
\put(124,1){$\varphi_{ab}$}
\put(103,23){$\varphi_{ca}$}
\put(139,23){$\varphi_{bc}$}

\end{picture}\\

\textbf{Definition} : To an arrow
$ \Gamma =\left( (\gamma _{i},\gamma _{i+1})\right) _{0\leq i\leq n-1} $
of $ \widehat{\Pi } $ corresponds the functor
$ (S_{\Gamma }:S_{\gamma _{0}}\rightarrow S_{\gamma _{n}}) $
defined by composition :\\
\begin{equation}
S_{\Gamma }=(S_{\gamma _{1}\gamma _{0}^{-1}})(S_{\gamma _{2}\gamma _{1}^{-1}})\cdot \cdot \cdot
(S_{\gamma _{n-1}\gamma _{n-2}^{-1}})(S_{\gamma _{n}\gamma _{n-1}^{-1}})
\end{equation}
\\
 This functor $ S_{\Gamma } $ will be called the sweeping functor (SF) associated
to the homotopy $ \Gamma  $.\\

\textbf{Remark} : To define a SF we need a succession of edge-paths and this
induces an orientation of the underlying 2-simplices. Another sequence of edge-paths
sweeping the same surface can define a different orientation of the 2-simplices
and thus a different SF, as illustrate the following pictures :\\

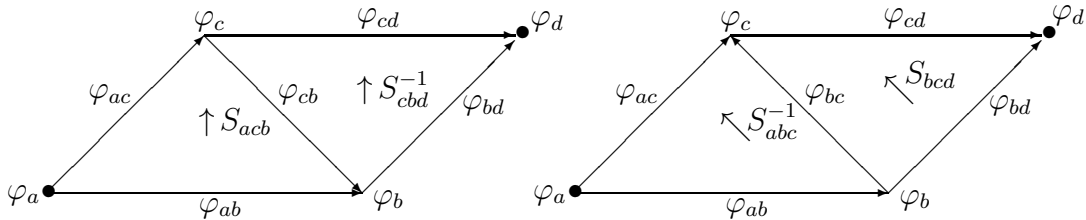
\begin{picture}(100,50)(-15,-10)

\put(0,0){\vector(1,0){60}}
\put(0,0){\vector(1,1){30}}
\put(30,30){\vector(1,-1){30}}
\put(60,0){\vector(1,1){29}}
\put(30,30){\vector(1,0){59}}

\put(-1,-1){$\bullet$}
\put(89,29){$\bullet$}

\put(-7.5,-1){$\varphi_{a}$}
\put(62,-2){$\varphi_{b}$}
\put(28,32){$\varphi_{c}$}
\put(92,32){$\varphi_{d}$}

\put(29,12){$\uparrow$}
\put(33,12){$S_{acb}$}

\put(29,-3.5){$\varphi_{ab}$}
\put(8,18){$\varphi_{ac}$}
\put(44,18){$\varphi_{cb}$}
\put(59,32.5){$\varphi_{cd}$}
\put(79,16){$\varphi_{bd}$}

\put(59,18){$\uparrow$}
\put(63,18){$S_{cbd}^{-1}$}

\put(100,0){\vector(1,0){60}}
\put(100,0){\vector(1,1){30}}
\put(160,0){\vector(-1,1){30}}
\put(160,0){\vector(1,1){29}}
\put(130,30){\vector(1,0){59}}

\put(99,-1){$\bullet$}
\put(189,29){$\bullet$}

\put(92.5,-1){$\varphi_{a}$}
\put(162,-2){$\varphi_{b}$}
\put(128,32){$\varphi_{c}$}
\put(191,33){$\varphi_{d}$}

\put(133,12){$S_{abc}^{-1}$}

\put(159,18){${\nwarrow}$}

\put(129,-3.5){$\varphi_{ab}$}
\put(108,18){$\varphi_{ac}$}
\put(144,18){$\varphi_{bc}$}
\put(159,32.5){$\varphi_{cd}$}
\put(179,16){$\varphi_{bd}$}

\put(163,21){$S_{bcd}$}

\put(128,11){$\nwarrow$}

\end{picture}\\

This process measures the curvature of the connection of $ S $ around a square
of $ \widehat{\Pi } $ which projects to a degenerated tetrahedron in $ P $
:\\
$$ S_{(ab,bd)}\stackrel{S_{acb}\otimes 1}{\longrightarrow }S_{(ac,cb,bd)}
\stackrel{1\otimes S_{cbd}^{-1}}{\longrightarrow }S_{(ac,cd)}\stackrel{S_{abc}\otimes 1}
{\longrightarrow }S_{(ab,bc,cd)}\stackrel{1\otimes S_{bcd}^{-1}}{\longrightarrow }S_{(ab,bd)} $$
 \vspace{1cm}

\section{A basic example}

\vspace{1cm}

Let's consider as an elementary illustration of the previous definitions the
case of a tetrahedron swept in two different ways determined by the following
pictures :\\

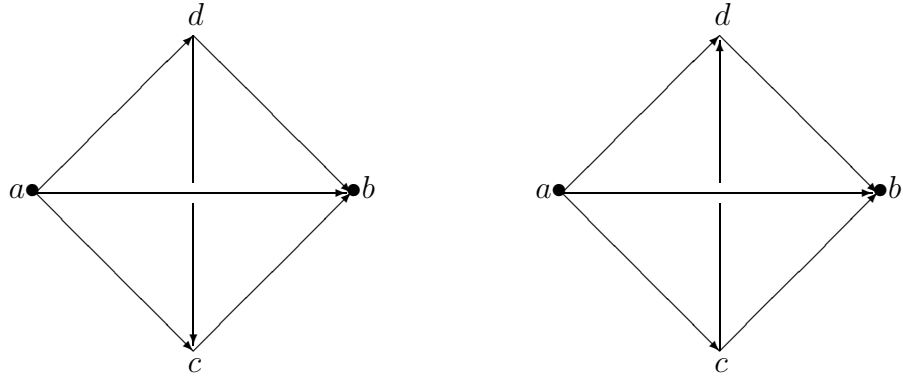
\begin{picture}(100,100)(-20,-50)

\put(10,0){\vector(1,0){59}}
\put(10,0){\vector(1,1){30}}
\put(10,0){\vector(1,-1){30}}
\put(40,30){\vector(1,-1){30}}
\put(40,-30){\vector(1,1){30}}
\put(40,30){\line(0,-1){28}}
\put(40,-2){\vector(0,-1){27}}

\put(8,-1){$\bullet$}
\put(69,-1){$\bullet$}
\put(5,-1){$a$}
\put(72,-1){$b$}
\put(39,32){$d$}
\put(39,-34){$c$}

\put(110,0){\vector(1,0){59}}
\put(110,0){\vector(1,1){30}}
\put(110,0){\vector(1,-1){30}}
\put(140,30){\vector(1,-1){30}}
\put(140,-30){\vector(1,1){30}}
\put(140,2){\vector(0,1){27}}
\put(140,-2){\line(0,-1){28}}

\put(108,-1){$\bullet$}
\put(169,-1){$\bullet$}
\put(105,-1){$a$}
\put(172,-1){$b$}
\put(139,32){$d$}
\put(139,-34){$c$}

\end{picture}\\

We start from the path $ (ac,cb) $ with the generic word
$ (\sigma _{ac},\sigma _{cb})=(x,y) $
whose letters form a connection above this initial path, we sweep the surface
path after path and we come back to the base path $ (ac,cb) $ with a final
word to be compared to the initial one. The succession of words in the first
case is : \\
\begin{eqnarray*}
(ac,cb) & \rightarrow  & (x,y)\\
(ab) & \rightarrow  & (xy\varphi _{acb}^{-1})\\
(ad,db) & \rightarrow  & (xy\varphi _{acb}^{-1},\varphi _{adb})\\
(ad,dc,cb) & \rightarrow  & (xy\varphi _{acb}^{-1},\varphi _{adb},\varphi _{dcb})\\
(ac,cb) & \rightarrow  & (xy\varphi _{acb}^{-1}\varphi _{adb}\varphi _{adc}^{-1},\varphi _{dcb})
\end{eqnarray*}
\\
 And in the second case, we have : \\
\begin{eqnarray*}
(ac,cb) & \rightarrow  & (x,y)\\
(ab) & \rightarrow  & (xy\varphi _{acb}^{-1})\\
(ad,db) & \rightarrow  & (xy\varphi _{acb}^{-1},\varphi _{adb})\\
(ac,cd,db) & \rightarrow  & (xy\varphi _{acb}^{-1},\varphi _{acd},\varphi _{adb})\\
(ac,cb) & \rightarrow  & (xy\varphi _{acb}^{-1},\varphi _{acd}\varphi _{adb}\varphi _{cdb}^{-1})
\end{eqnarray*}
\\
 These two sweeping (or pasting) schemes are related by a flat tetrahedron, as described
in the previous  remark. In order to compare the final words to the initial one,
we can make a gauge transformation on the final connection so that we obtain
a word of the form $ (xg_{1},yg_{2}) $. If we compare this final word to
the initial one, the insertion of the $ g_{i} $'s defines a notion of 2-holonomy.
This operation depends of course on the sweeping scheme. In order to obtain
an invariant notion of 2-holonomy, we must lift unparametrized surfaces in a
fibered category having "more symmetry". \\

\section{Abelianisation}

\vspace{1cm}

In the previous sections, we have worked on the space $ \widehat{\Pi } $
of paths in $ P $. The sweeping functors defined above depend on the sequence
of paths $ \gamma _{i} $ which sweep the surface in $ X $. 
How can we get rid of this dependence and obtain invariant sweeping functors 
which depend only on the swept surface ? The pasting theorem \cite{pasting} 
states that the composition of a family of composable 2-arrows in any 2-category 
depends only on the planar oriented graph (or pasting scheme) they define, 
and not on the order of composition. Each pasting scheme is a discrete flow on 
the swept surface and the evolution of a path is a succession of local homotopies. 
Two distant homotopies must be represented algebraically by operations 
which commute. This is indeed the case when we represent these 
motions by insertions of letters in a word. This commutativity is a natural consequence 
of locality, but there is another commutativity which appears if we are not careful 
with the objects on which the infinitesimal homotopies act. 
A mistake is to put the same fiber, say $G_1$, above each trivial loop and to connect them by the identity functor, believing that such a trivial gerbe on the path space can have interesting holonomies on surfaces since only the 2-cells carry the "fluxes" which are our physical degrees of freedom. 
But if we suppose that the connective structure is trivial, 
$\phi_{ab}= \mathrm{Id} : G_1 \to G_1$ is the trivial $G$-bitorsor for all $ a,b \in X^0$. 
For each oriented triangle $ (acb) $, for all $ x,y\in O(\varphi _{a}) $ and for
any arrow $ (u:x\rightarrow y) $, the naturality of
$ (\varphi _{acb}:\varphi _{ab}\rightarrow \varphi _{ac}\varphi _{cb}) $
means the following equivariance condition :\\

\begin{equation}
\varphi _{ab}(u)\varphi _{acb}(y)=\varphi _{acb}(x)\varphi _{ac}\varphi _{cb}(u)
\end{equation}
\\

The relation (6) implies that for $ x=y $, $ \varphi _{acb}(x) $ 
commutes with every $ u\in \mathrm{Aut}(x)\simeq G $
thus $ \varphi _{acb} $ lands in $ Z(G) $, the center of $ G $. 
This pitfall is related to the argument used by Eckman and Hilton to prove the 
commutativity of the second homotopy group of a topological space \cite{BD}. 

In fact, the most general sweeping functors transport the sections of the fibered 
category $ \varphi _{u} $, which lives above the initial path $ u $, to a section
of $ \varphi _{v} $ above the final path $ v $, the edge-paths $ u $
and $ v $ being $ X^{2} $-homotopic. These sections are words of objects.
By sweeping each oriented triangle, one inserts a group element in these words
or regroups neighbouring letters, as indicated by (1-4). This process depends
on the order and orientation of the elementary 2-arrows, as shown in the last
remark of the previous section. In a smooth limit, to be precisely defined,
we expect the sweeping functors to depend on the parametrisation of the swept
surface and to form a representation space for the group of diffeomorphisms
of this surface.\vspace{1cm}

\section{Conclusion and Perspectives}

\vspace{1.0cm}

Let's summarise our work. We have considered a polyhedron $ P $ as a base
space. We have defined its simplicial groupoid $ \Pi  $ as the category
which has the vertices as objects and the edge-paths as arrows. Having fixed a
locally compact topological group $ G $, we have defined the simplicial $ G $-bundles
with connection over $ P $ as the functors from $ \Pi  $ to the groupoid
$ G_{1} $ of (right) $ G $-torsors. The fibers are associated to the vertices
of $ P $ and are connected by the morphisms associated to the edges of $ P $.
Then, we have extended this definition to two dimensions, with the groupoid
$ \widehat{\Pi } $. The objects of $ \widehat{\Pi } $ are the edge-paths
of $ P $ and the arrows of $ \widehat{\Pi } $ are obtained by sweeping
a finite number of oriented triangles in a consistent way from an initial path
to a final one (with the same boundary points). A fibered category over $ \widehat{\Pi } $
is then constructed, which associates to each path the category of sections
of a stack in groupoids over $ P $, restricted to this path. The connection
of this fibered category of sections associates to each combinatorial homotopy
$ \Gamma  $ between two edge-paths a functor $ S_{\Gamma } $ which transports
the sections of fibered categories along the surface. On a smooth surface 
$ \Sigma  $ embedded in $ X $, the 2-cells of a pasting diagram define a simplicial 
approximation of a smooth flow and $ \mathrm{Diff}(\Sigma ) $ could act on the space of
general sweeping functors, thus defining a representation of this group of diffeomorphisms.

Perhaps the main lesson of our approach is that the discretization of geometric,
set-theoretic objects leads naturally to a description in terms of categories
and higher algebraic structures. This functorial approach seems adapted to gauge
theories and string theories. Indeed, when the energy concentrates on thin flux
tubes, on loops or on membranes, these new excitations carry information on
higher dimensional cells and must be lifted into a suitable fibered $ n $-category.
A study of these spaces could shed a new light on the duality relations.

\vspace{1cm}

\textbf{Acknowledgements} : I wish to thank Daniel Bennequin for his constant
help and encouragement, Lawrence Breen who gave me Ariane's thread of functoriality,
and Jean-Luc Brylinski for his critics of my early work. I am also grateful to James Stasheff
for a careful reading of the preprint.\vspace{1cm}

\textbf{Note added} : During the publication of the present paper, Lawrence Breen informed me of his joint work with William Messing about the differential geometry of gerbes \cite{BM}. They found new identities satisfied by the connection, 2-curvature and 3-curvature, which may have a simple interpretation in our functorial context (work in progress ...).

\end{document}